\documentclass[preprint,prd,aps,pdftex,showpacs,preprintnumbers,amsmath,amssymb]{revtex4-1}
\usepackage{graphics,epsfig,subfigure}
\usepackage{diagbox}
\usepackage[usenames]{color}
\usepackage{color}
\usepackage{graphicx}
\usepackage{amsfonts}
\usepackage{indentfirst}
\usepackage{booktabs}
\usepackage{hyperref}
\hypersetup{hypertex=ture,backref=true,colorlinks=ture,linkcolor=blue,anchorcolor=blue,citecolor=blue}
\usepackage{float}
\usepackage{multirow}

\begin{document}

\title{Compact Stars Sourced by Perfect Fluid Dark Matter Halos}

\author{Yuan Yue$^{1}$}
\author{Yong-Qiang Wang$^{2}$}\email{yqwang@lzu.edu.cn, corresponding author}
\affiliation{$^{1}$College of Mathematics and Computer Science, Northwest Minzu University, Lanzhou, 730030, China\\
$^{2}$School of Physical Science and Technology, Lanzhou University, Lanzhou 730000, China}

\begin{abstract}
Recent studies have shown that dark matter halos can support regular black holes or compact stars by assuming an anisotropic energy-momentum tensor. In this paper, we extend the analysis to the dark matter halo as an isotropic perfect fluid. By employing galactic dark matter profiles—specifically the Einasto and Dehnen models—as the mass-energy density source, we numerically solve the Einstein field equations and find a class of non-singular, horizonless compact star solutions. Moreover, these configurations remain stable against axial perturbations while satisfying the dominant  energy condition.
\end{abstract}

\maketitle

\section{Introduction}
Dark matter halos serve as the primary gravitational scaffolds for the formation and evolution of galaxies and clusters \cite{White1978,Springel2008}. Their existence is supported by robust observational evidence, most notably the flattened galactic rotation curves that deviate from Newtonian predictions \cite{Rubin1970,Sofue2001}. Additionally, gravitational lensing enables the direct mapping of total mass-energy density independent of luminosity \cite{Bartelmann2001, Massey2007}, while indirect searches for gamma-ray signatures provide complementary constraints on halo distributions \cite{Bertone2005}. Together, these multi-messenger probes establish dark matter halos as a fundamental gravitational source in astrophysical modeling.

Recently, the central regions of dark matter halos have attracted significant interest, particularly as potential matter sources for regular black hole (RBH) geometries. Early studies explored this in (2+1)-dimensional gravity  \cite{Sajadi:2023ybm} or through multi-polytropic equations of state \cite{Sajadi:2025prp}. More recently, it was shown that DM halos described by realistic profiles, such as the Einasto or Dehnen models, can support RBH spacetimes \cite{Kar:2025phe,Konoplya:2025ect}. 
These developments have prompted comprehensive studies on their observational signatures, including quasinormal modes (QNMs), grey-body factors, and absorption cross-sections across various fields \cite{Lutfuoglu:2025grav, Bolokhov:2025qnm, Saka:2025, Malik:2025}. 
However, these non-singular black hole solutions typically require a stringent constraint: the radial pressure must be the negative of the energy density ($P_r = -\rho$).  Recent analyses \cite{
Bolokhov:2025crit}  point out that many existing solutions fail to consistently satisfy the Einstein equations for their intended matter distributions.
What occurs if this stringent constraint is relaxed? Recent work \cite{Yue:2026evf} has generalized the anisotropic energy-momentum tensor by relaxing the $P_r = -\rho$ condition and shown that while RBHs represent a unique special case, a broader class of pressure-density relations leads to horizonless compact stars. Under specific parameter limits, these objects approach a ``frozen state" that mimics black hole features without possessing an event horizon, which also were found in both boson stars \cite{Wang:2023tdz,Yue:2023sep,Chicaiza-Medina:2025wul} and neutron stars \cite{Tan:2025jcg} within the Bardeen and Hayward models.

Despite these advancements, the reliance on anisotropic pressure—where radial and tangential pressures differ ($P_r \neq P_\perp$)—remains a noteworthy issue. 
Observations allow for a direct inference of energy density (mass distribution), yet pressure—being a dynamical property—requires derivation from theoretical models and underlying physical assumptions.
A natural  question arises: Can a  isotropic perfect fluid ($P_r = P_\perp$) serve as the source for such dark matter-embedded geometries? Furthermore, can such a source sustain a stable configuration that avoids the formation of singularities?

In this paper, we extend the investigation of dark matter halos by modeling them as isotropic perfect fluids. By adopting the Einasto and Dehnen galactic profiles as the  energy density sources, we numerically  solve the Einstein field equations with a purely isotropic pressure to obtain a novel class of non-singular, horizonless compact star solutions. Our analysis further demonstrates that these configurations exhibit robust stability against axial perturbations while satisfying the dominant  energy condition (DEC).

The structure of this paper is as follows: Sec. II introduces  the model within the framework of an isotropic perfect fluid dark matter halos. Sec. III presents numerical  solutions for compact stars under various dark matter halo density profiles, and investigates the stability of these backgrounds under axial perturbations. Finally, we summarize our conclusions in Sec. IV.

\section{MODEL SETUP}
We consider the Einstein field equations in the presence of a dark matter source:
\begin{equation}\label{eom1_lower}
    G_{\mu\nu} = R_{\mu\nu} - \frac{1}{2} g_{\mu\nu} R = 8\pi T_{\mu\nu},
\end{equation}
where $T_{\mu\nu}$ is the energy-momentum tensor representing an isotropic perfect fluid:
\begin{equation}
    T_{\mu\nu} = (\rho + P) u_\mu u_\nu + P g_{\mu\nu}. \label{eq:EnergyMomentum_lower}
\end{equation}
Here, $\rho$ and $P$ denote the energy density and isotropic pressure, respectively, while $u_\mu$ is the four-velocity of the fluid. To describe a static, spherically symmetric spacetime, we adopt the following metric:
\begin{equation}\label{equ10}
	ds^2 = -N(r)\sigma^2(r)dt^2 + \frac{dr^2}{N(r)} + r^2 d\Omega^2,
\end{equation}
where $d\Omega^2 = d\theta^2 + \sin^2\theta d\varphi^2$ is the metric on a unit two-sphere. In the co-moving frame, the four-velocity is given by $u_\mu = (-\sqrt{N}\sigma, 0, 0, 0)$.
Under this metric ansatz, the non-vanishing components of the Einstein equations yield:
\begin{align}
  G^t{}_t &=   -8\pi \rho = \frac{1}{r^2} \left( r N' + N - 1 \right), \label{eq:field1} \\
  G^r{}_r &=   8\pi P = \frac{1}{r^2} \left[ N \left( \frac{2r\sigma'}{\sigma} + 1 \right) + r N' - 1 \right], \label{eq:field2} \\
  G^\theta{}_\theta = G^\phi{}_\phi &=   8\pi P = \frac{N''}{2} + \frac{N'}{r} + \frac{3 N' \sigma'}{2 \sigma} + \frac{N \sigma''}{\sigma} + \frac{N \sigma'}{r \sigma}, \label{eq:field3}
\end{align}
where the prime denotes a derivative with respect to the radial coordinate $r$. The local conservation of energy-momentum, $\nabla_\mu T^{\mu\nu} = 0$, for a static and spherically symmetric metric, the only non-trivial component of the conservation equation is the radial one ($\nu = r$), leads to the Tolman-Oppenheimer-Volkoff (TOV) equation for a perfect fluid:
\begin{equation}
    P' + \frac{1}{2} \left( \frac{N'}{N} + \frac{2\sigma'}{\sigma} \right) (\rho + P)= 0. \label{eq:TOV}
\end{equation}
Among these equations  \ref{eq:field1}, \ref{eq:field2}, \ref{eq:field3}, and \ref{eq:TOV}, only three are independent;  the ${}^\theta{}_\theta$ component of the Einstein equations  can be derived from the aforementioned equations. To solve the above equations, one must specify a dark matter energy density profile $\rho(r)$. It is worth noting that under the specific equation of state $\rho + P = 0$, the TOV equation becomes trivial ($P' = 0$), and the system recovers AdS black hole solutions. For the general case, the functions $N(r)$, $\sigma(r)$, and $P(r)$ are determined by solving the coupled system \eqref{eq:field1}, \eqref{eq:field2}, and \eqref{eq:TOV}. These solutions are subject to the following boundary conditions:
\begin{align}
    N(0) &= 1, & N(r) &\xrightarrow{r\to\infty} 1-\frac{2M}{r},  \nonumber\\
    \sigma(r) &\xrightarrow{r\to\infty} 1, & P(r) &\xrightarrow{r\to\infty} 0. \label{equ20}
\end{align}
These requirements ensure that the configurations are regular at the origin and approach asymptotic flatness at infinity.

\section{Numerical results}\label{sec3}

To integrate the coupled system of non-linear ordinary differential equations, namely the field equations \eqref{eq:field1}--\eqref{eq:field2} and the TOV equation \eqref{eq:TOV}, we implement a numerical scheme subject to the boundary conditions specified in \eqref{equ20}. Given the semi-infinite nature of the radial coordinate $r \in [0, \infty)$, it is computationally advantageous to compactify the domain. We therefore introduce a dimensionless coordinate $x$ defined by the map\begin{equation}x = \frac{r}{1+r} , \label{eq:compact_coord}\end{equation}which maps the spatial infinity to the boundary $x=1$. The resulting system of non-linear algebraic equations, obtained after discretization, is solved via an iterative Newton-Raphson algorithm. To ensure the robustness and high fidelity of our numerical results, we impose a stringent convergence criterion, maintaining a relative error tolerance of better than $10^{-8}$ across the entire integration grid.

In the following analysis, we consider two distinct scenarios for the mass distribution within these halos:
\begin{itemize}
    \item \textbf{Einasto-type density \cite{Einasto1965,EinastoHaud1989,Retana2012,Konoplya:2025} :}
   \begin{equation}
 \rho(r) = \rho_{0} \exp \left[ -\left( \frac{r}{h} \right)^{1/n} \right], \quad n > 0.
\end{equation}
    \item \textbf{Dehnen-type density \cite{Dehnen:1993uh,Taylor:2002zd}:} 
    \begin{equation}\label{Dehnendensity}
\rho(r)=\rho_0 \left(\frac{r}{a}\right)^{-\alpha} \left(1+\frac{r^k}{a^k}\right)^{-(\gamma-\alpha)/k}.
\end{equation}
    
\end{itemize}

\begin{figure}
  \includegraphics[width=8.2cm]{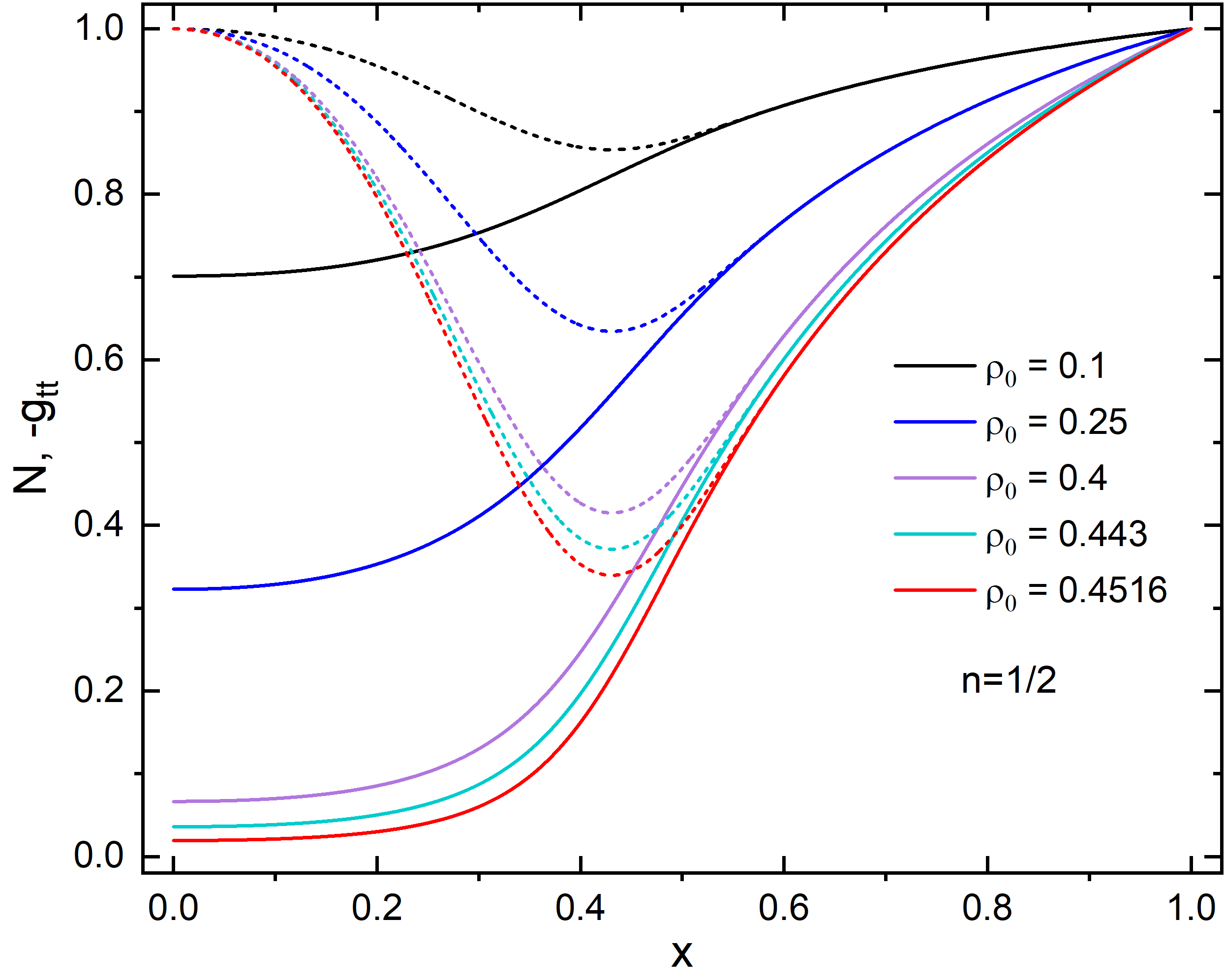}
 \includegraphics[width=8.1cm]{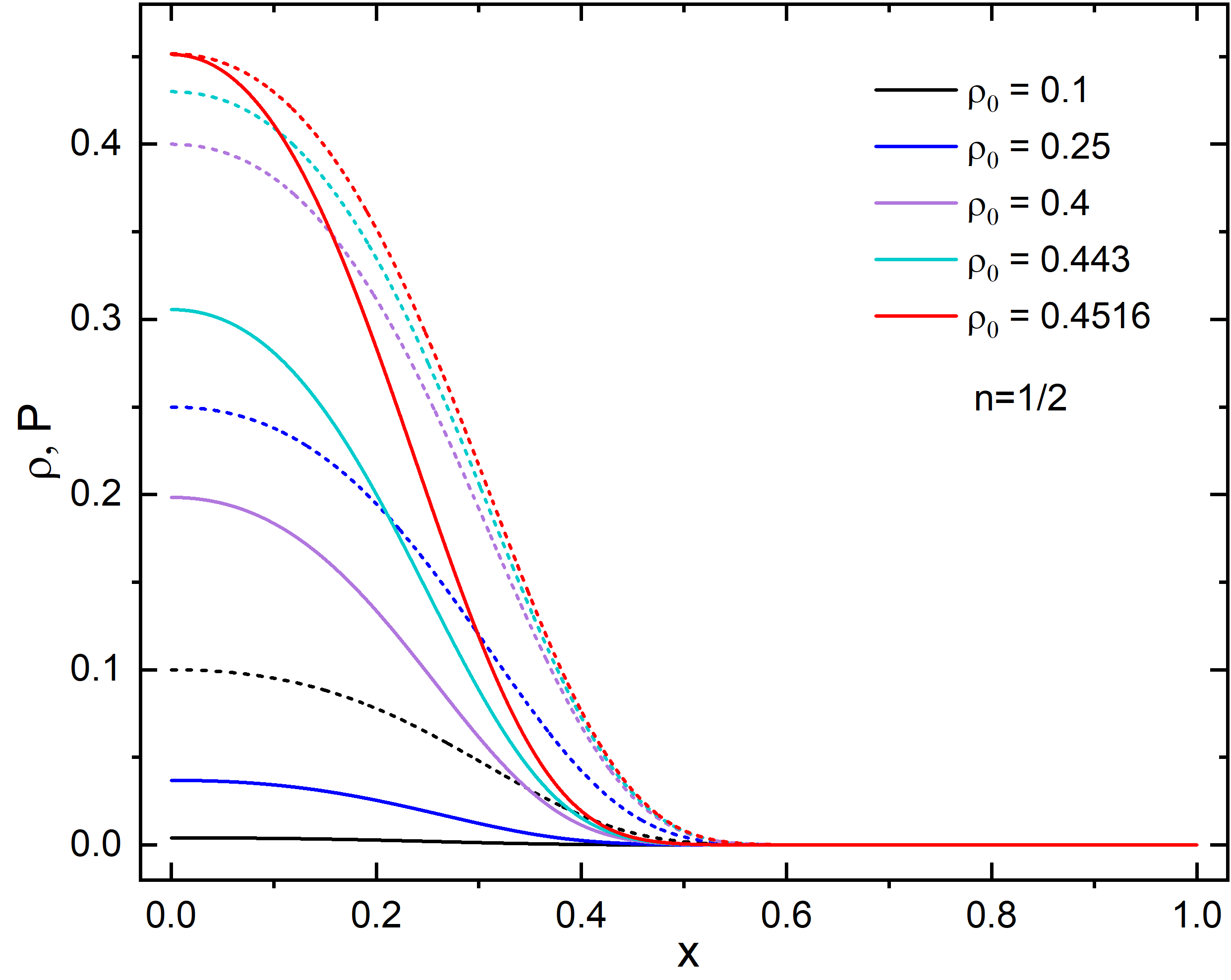}
  \begin{center}
      \includegraphics[width=8.1cm]{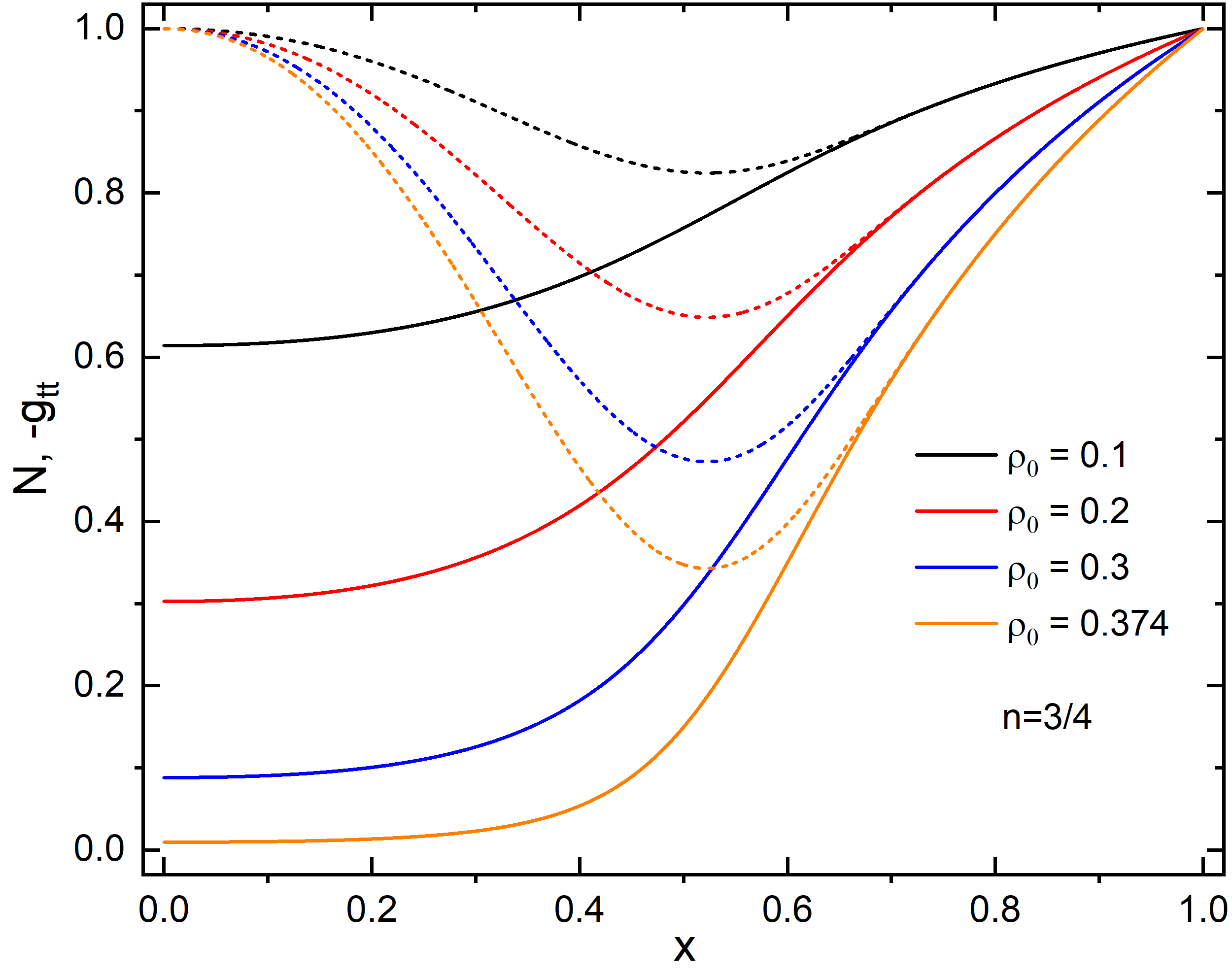}    \includegraphics[width=8.1cm]{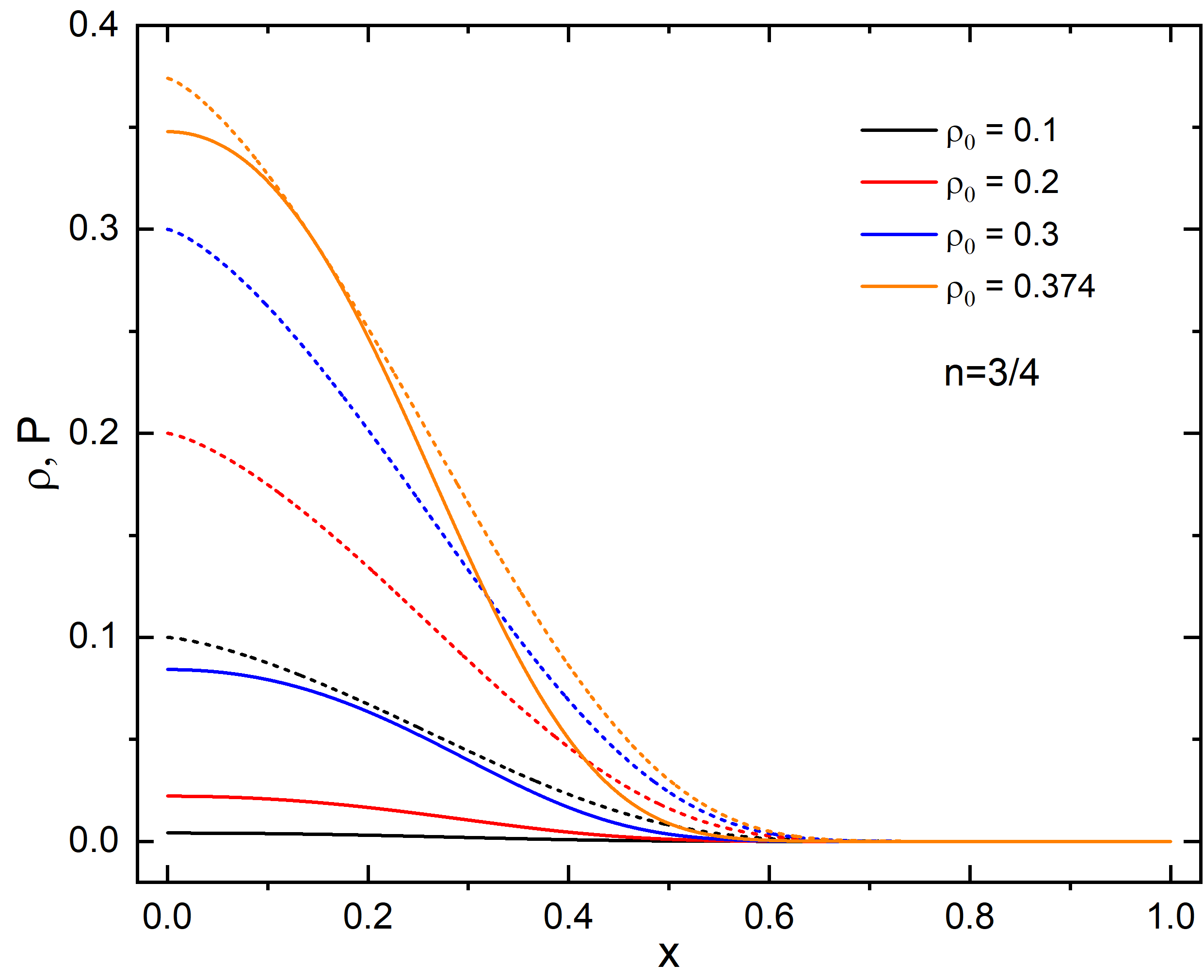}   \includegraphics[width=8.1cm]{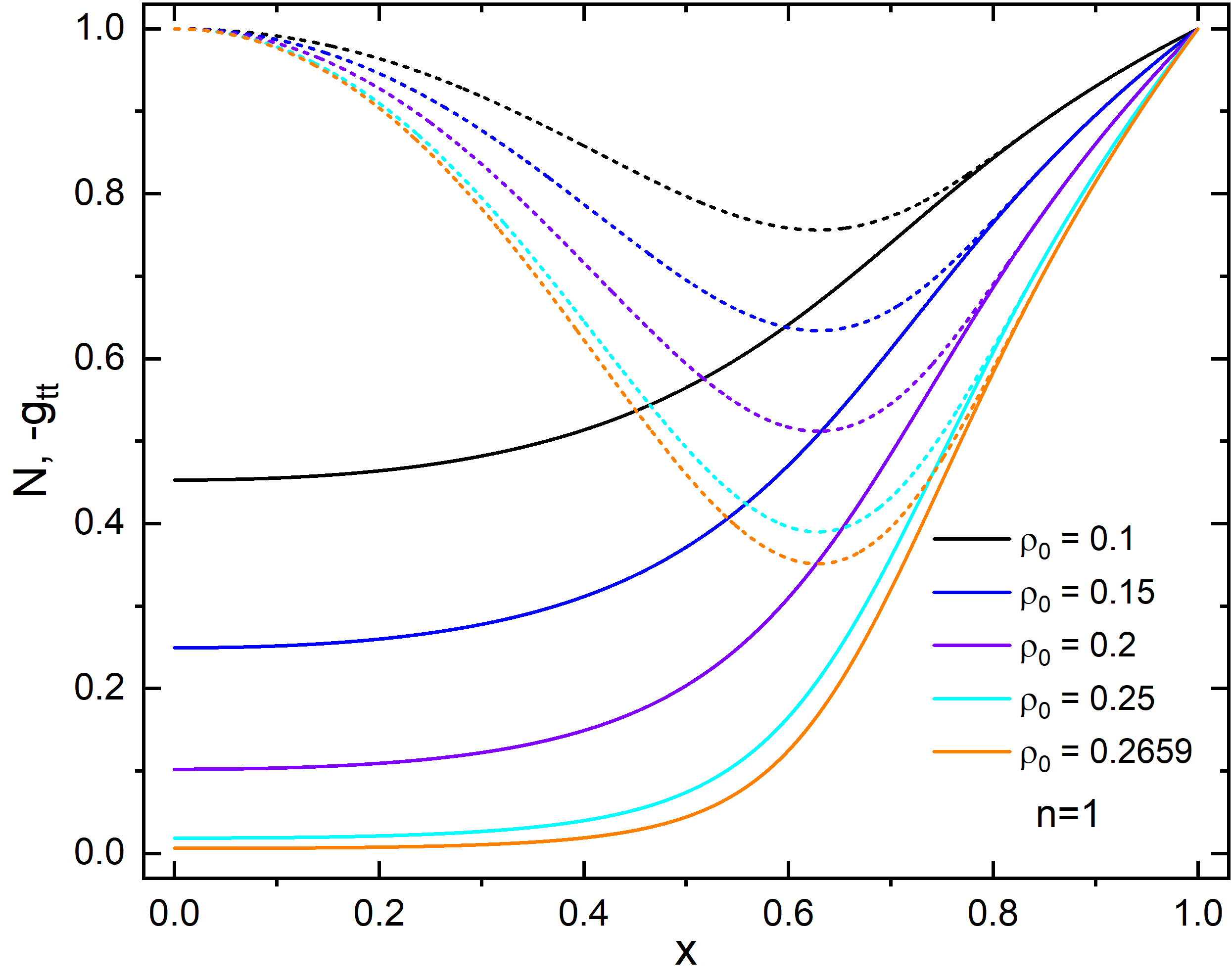}
      \includegraphics[width=8.1cm]{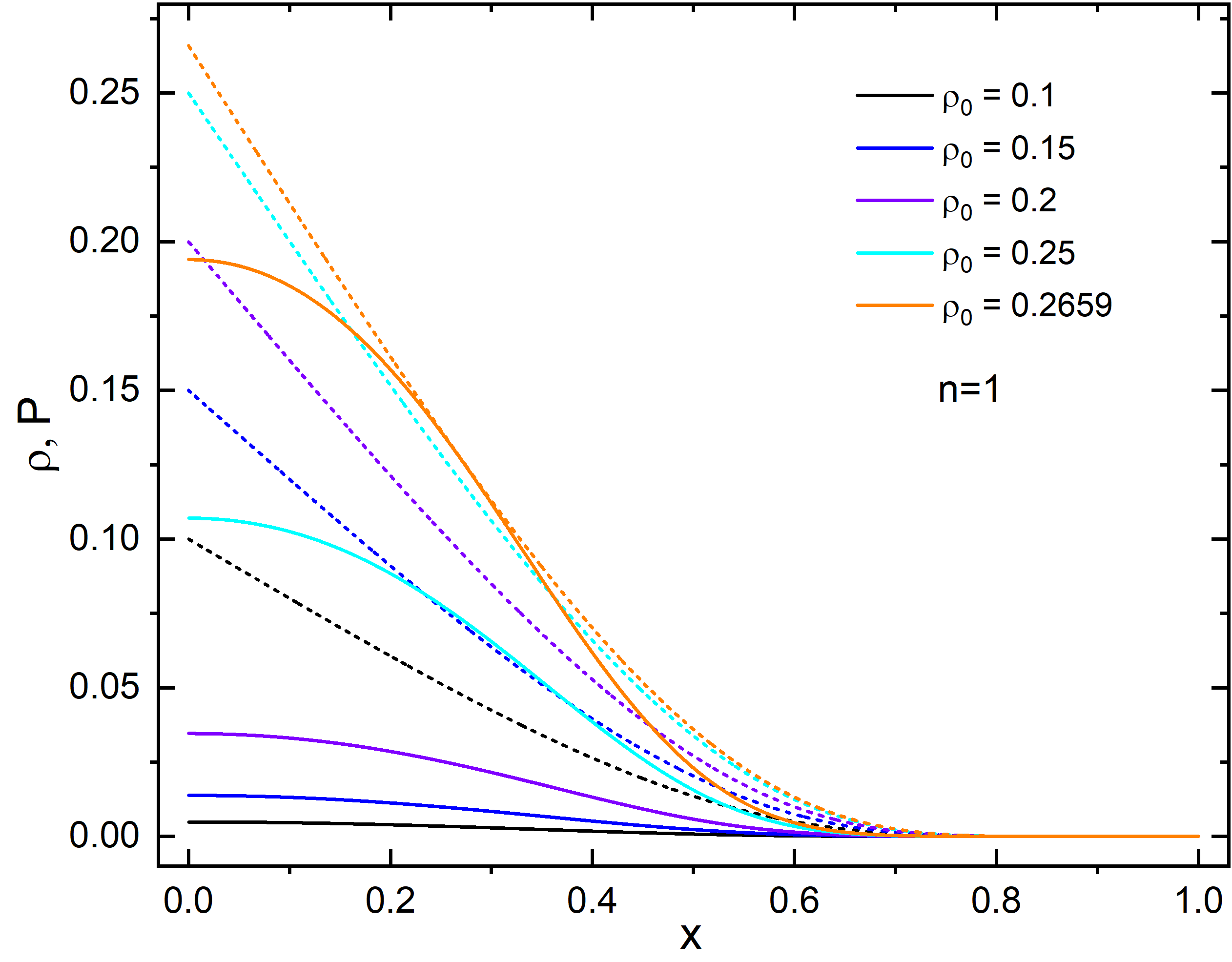}
  \end{center}
  \caption{Radial profiles of the functions  $N$,   $-g_{tt}$, $\rho$, and $P$ for an Einasto-type density distribution. The curves correspond to different values of the central density $\rho_0$, with  $n=0.5$, $0.75$, and $1$, respectively.
  }\label{phase}
\end{figure}

Furthermore, our subsequent calculations reveal that the isotropic pressure $P$ remains positive-definite throughout the domain, ensuring that the weak energy condition (WEC) is always satisfied. During the investigation of the parameter space, certain configurations are found to violate the dominant energy condition  $\rho \ge P$. To ensure the  causality and stability of the solution, we restrict our analysis to the region of parameter space where the solutions remain strictly consistent with the DEC.

\subsection{Einasto-type density}
In Fig. \ref{phase},  we present the spatial distributions of functions $N$ (dashed lines),   $-g_{tt}$ (solid lines), $\rho$ (dashed lines) , and $P$ (solid lines)   for different values of $\rho_0$ with  $n=0.5$, $0.75$, and $1$, respectively. 
It is observed that as $\rho_0$ increases, the minimum value of $-g_{tt}$ and  $N$ decrease significantly. 
Furthermore, for a fixed value of $\rho_0$, the minimum of these two functions decreases monotonically as the  $n$ increases.   Meanwhile, the numerical results indicate that the magnitude of the isotropic pressure $P$ remains positive-definite throughout the domain and increases monotonically with the growth of $\rho_0$.

A crucial feature revealed by the numerical analysis is the existence of a critical threshold for the central density, denoted as $\rho_{0,\text{c}}$. For the specific cases considered, we find $\rho_{0,\text{c}} = 0.4516, 0.374,$ and $0.2659$, corresponding to $n = 0.5, 0.75,$ and $1$, respectively.
At this threshold, the profiles of $\rho$ and $P$ intersect at a specific radial point. In the regime where $\rho_0 > \rho_{0,\text{c}}$, the pressure $P$ exceeds the energy density $\rho$ at certain locations, signaling a violation of the dominant energy condition. Conversely, for $\rho_0 \le \rho_{0,\text{c}}$, the condition $P \le \rho$ is satisfied globally, thereby ensuring the stability and physical viability of the halo configuration under the DEC.
The critical threshold for the central density, $\rho_{0,\text{c}}$, decreases with the increase of the $n$. 

It is noteworthy that for the case of $n=1/2$, the violation of the dominant energy condition originates at the center of the compact star solutions. In contrast, for $n=0.75$ and $n=1$, the violation does not first occur at the center; instead, the radial position of this violation shifts outward as the $n$ increases.

\subsection{Dehnen-type density}
\begin{figure}[]
  \begin{center}
  \includegraphics[width=8.1cm]{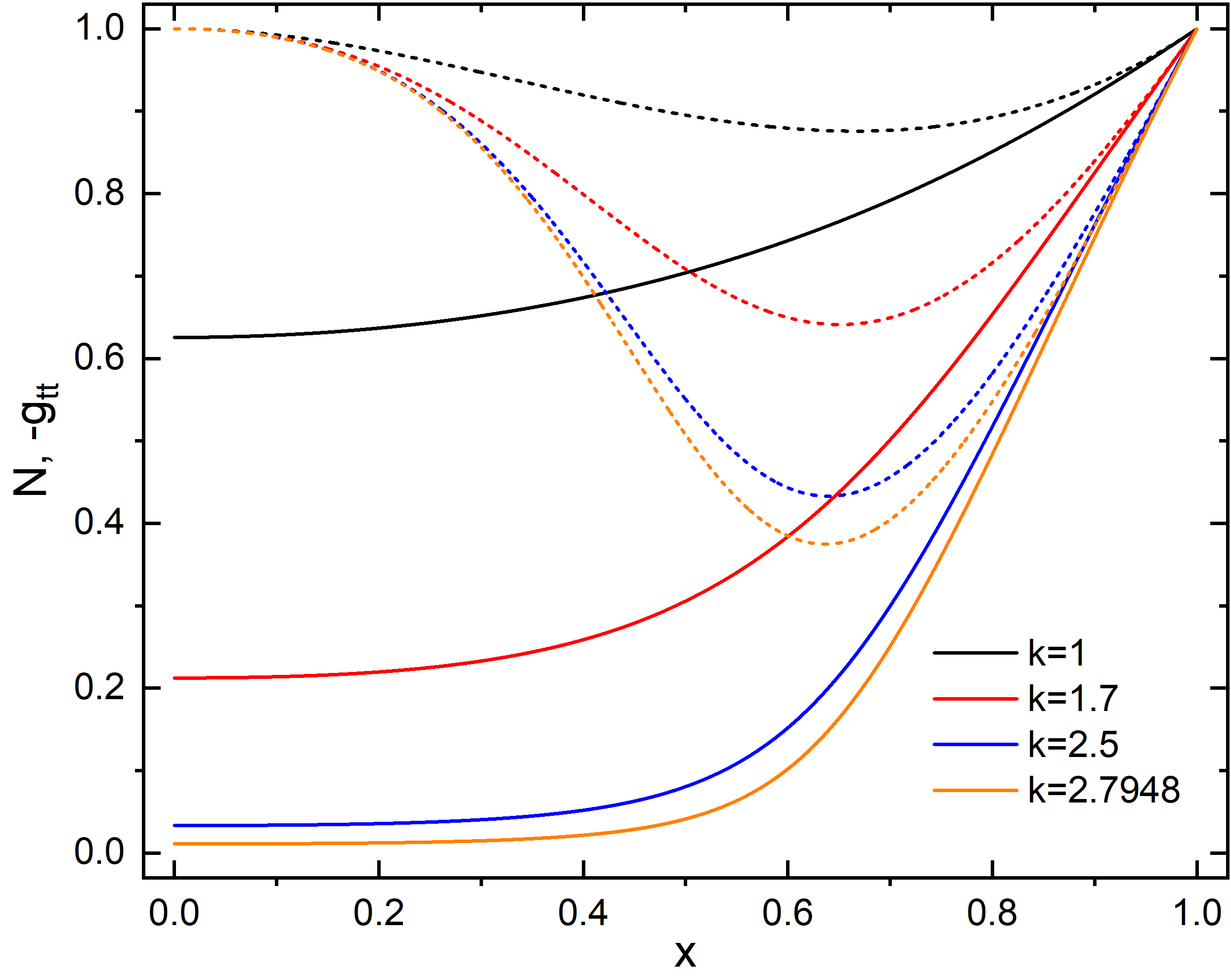}
   \includegraphics[width=8.1cm]{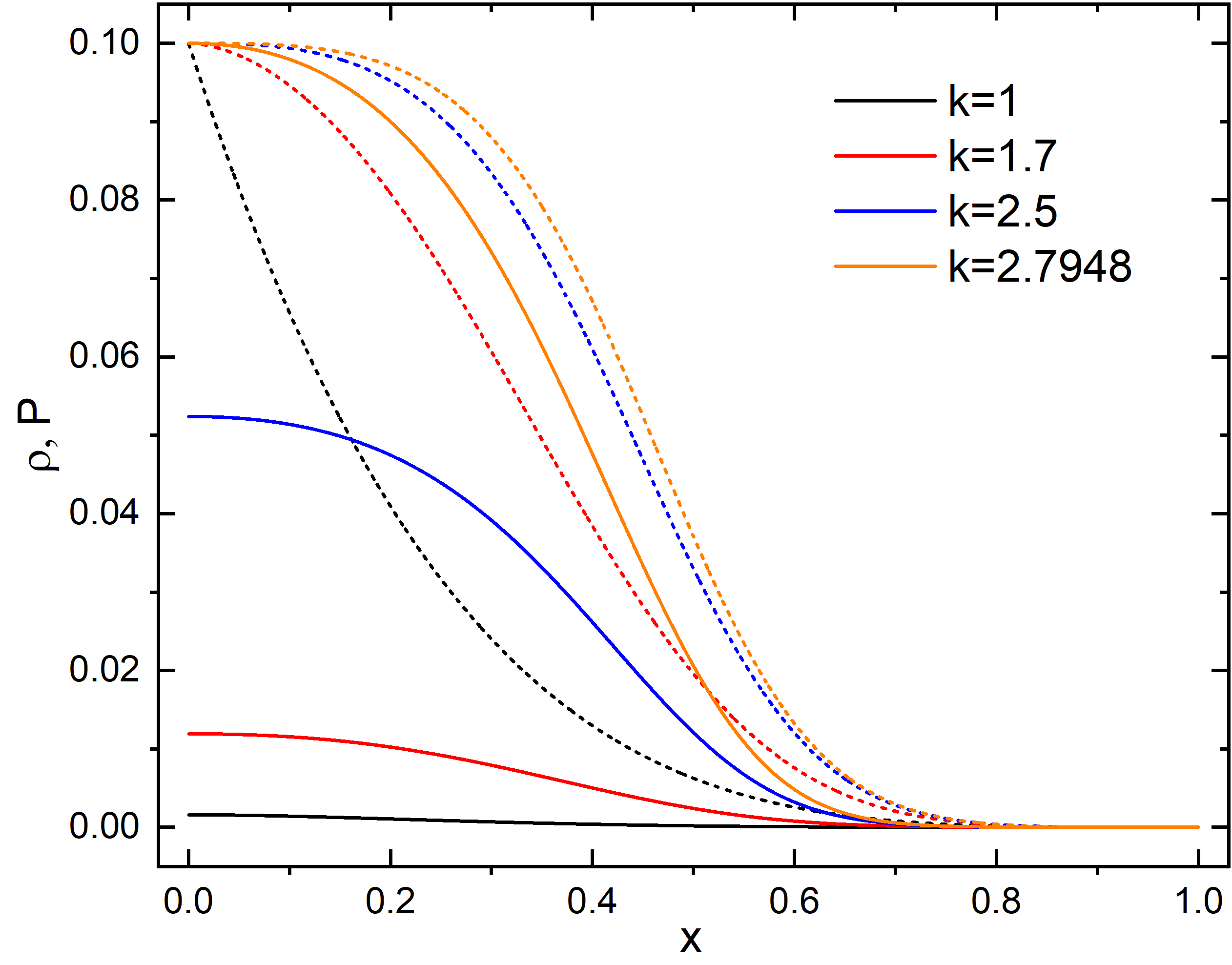}
  \end{center}
  \caption{Radial profiles of the functions  $N$,   $-g_{tt}$, $\rho$, and $P$ for an Dehnen-type density distribution. The curves correspond to different values of the  $k$, with $\rho_0=0.1$, $a=1$, $\gamma=4$, $\alpha=0$.
  }\label{gttsigpru1}
\end{figure}
 Fig. ~\ref{gttsigpru1} displays the spatial distributions of the  functions $N$ (dashed lines),   $-g_{tt}$ (solid lines), $\rho$ (dashed lines) , and $P$ (solid lines)   for different values of $k$. It is found that as $k$ increases, the minimum value of $-g_{tt}$ and  $N$ decrease. 
   Meanwhile, the numerical results indicate that the magnitude of the isotropic pressure $P$ remains positive-definite throughout the domain and increases monotonically with the growth of $k$.  There exists a critical threshold for $k_c = 2.7948$, beyond which the dominant energy condition is violated.
Moreover, the violation of the dominant energy condition originates at the center of the compact star solutions.

\begin{figure}[]
  \begin{center}
  \includegraphics[width=8.1cm]{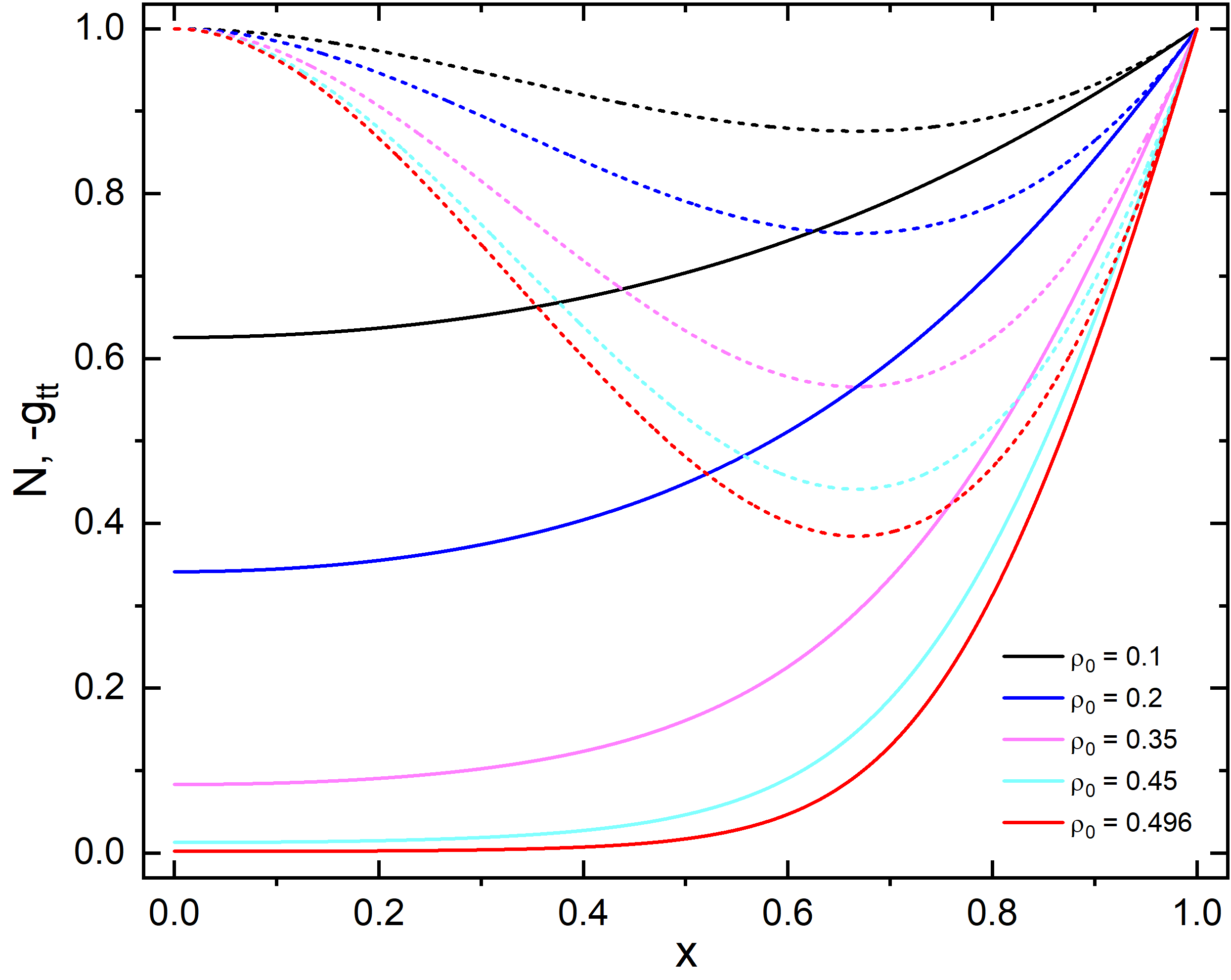}
   \includegraphics[width=8.1cm]{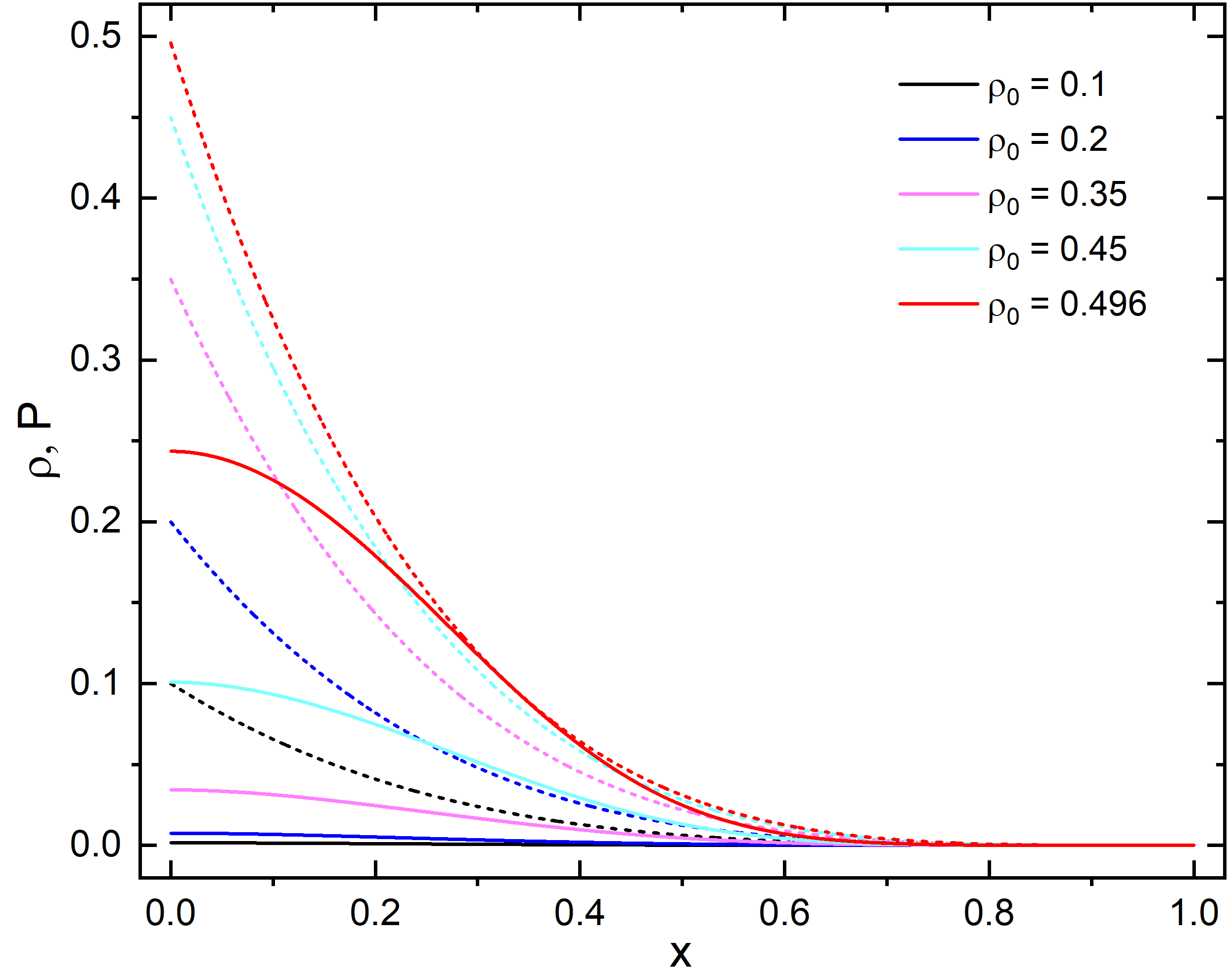}
  \end{center}
  \caption{Radial profiles of the functions  $N$,   $-g_{tt}$, $\rho$, and $P$ for an Dehnen-type density distribution. The curves correspond to different values of the central density $\rho_0$, with $a=1$, $k=1$, $\gamma=4$, $\alpha=0$.
  }\label{linjie}
\end{figure}

\begin{figure}[]
  \begin{center}
     \includegraphics[width=8.1cm]{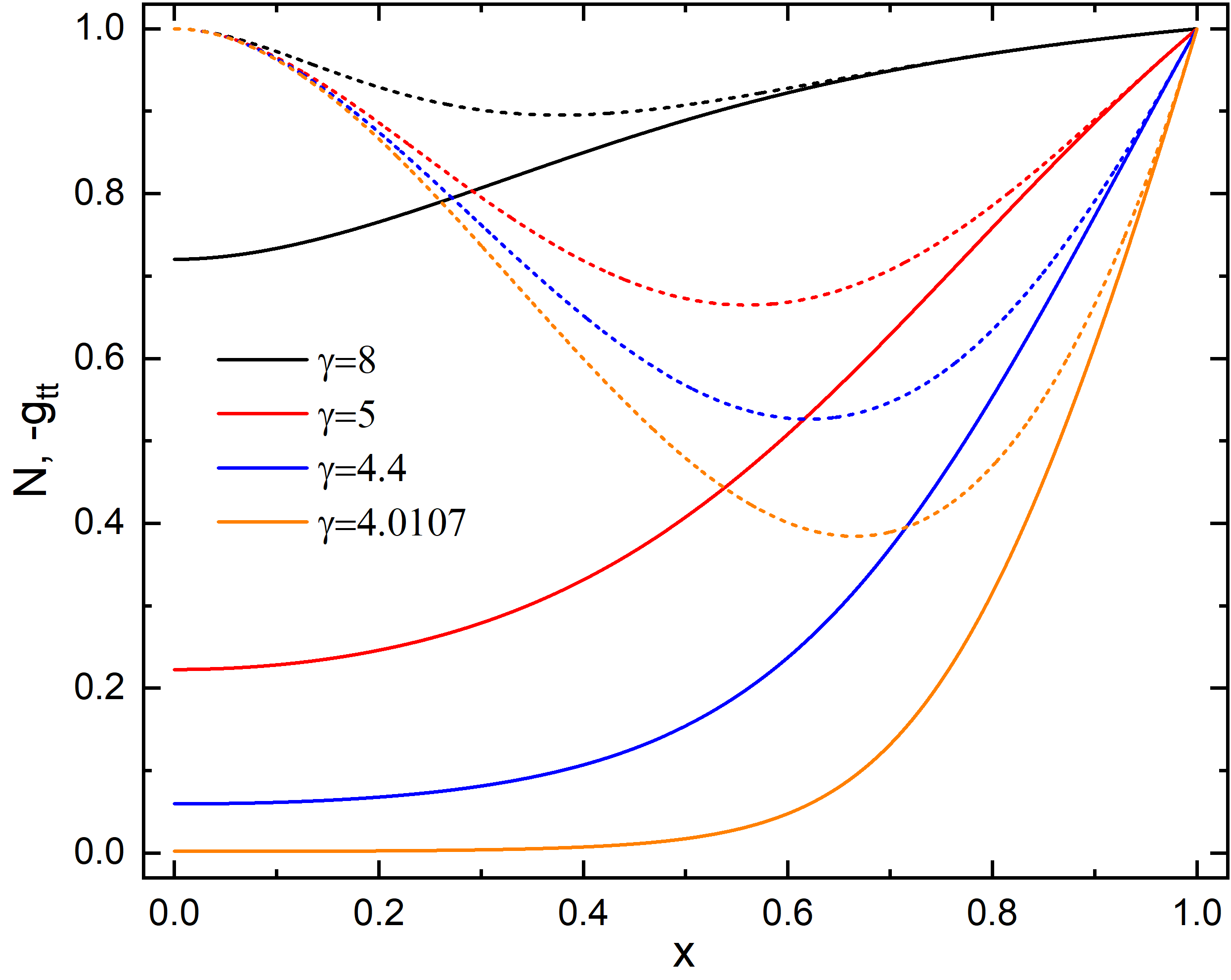}
   \includegraphics[width=8.1cm]{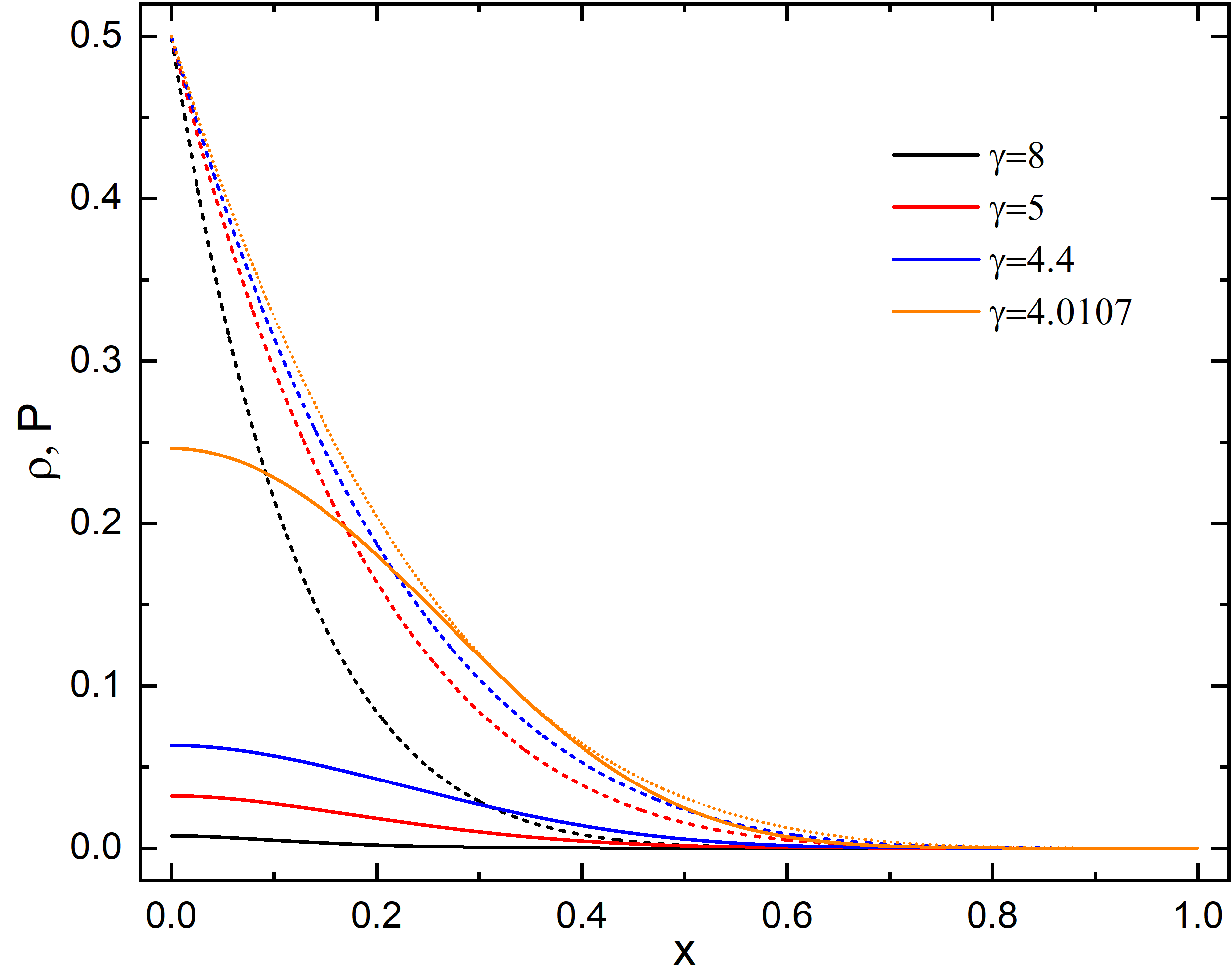}
  \end{center}
  \caption{Radial profiles of the functions  $N$,   $-g_{tt}$, $\rho$, and $P$ for an Dehnen-type density distribution. The curves correspond to different values of the  $\gamma$, with $\rho_0=0.5$, $a=1$, $k=1$, $\alpha=0$.
  }\label{phaseca2}
\end{figure}

Analogously, the spatial distributions of these functions for varying central density $\rho_0$ and parameter $\gamma$ are presented in Figs. \ref{linjie} and \ref{phaseca2}. The system exhibits a qualitative behavior consistent with the $k$-variation observed previously. Specifically, as $\rho_0$ or $\gamma$ increases, the minimum values of the metric functions further decrease, while the pressure magnitude $P$ is enhanced. The physical viability of these solutions is likewise bounded by critical limits, specifically $\rho_{0,c}=0.496$ and $\gamma_{c}=4.0107$. Beyond these thresholds, the DEC is no longer satisfied, marking the transition to non-physical regimes.

\subsection{Stability analysis of axial perturbations}
In the framework of the Regge-Wheeler formalism, the axial gravitational perturbations can be decoupled into a master wave-like equation \cite{Chakraborty:2024gcr}:
\begin{equation}
    \frac{d^2 \Psi}{dr_*^2} + \left( \omega^2 - V(r) \right) \Psi = 0, \quad d r_{*} \equiv \frac{d r}{f(r)},
\end{equation}
where $r_*$ is the tortoise coordinate. For both "up" and "down" perturbation sectors, the corresponding effective potentials are found to be identical for anisotropic perfect fluid and are given by:
\begin{equation}
\begin{aligned}
V^{(\text{up})}(r) &= V^{(\text{down})}(r) 
&= N \sigma^2 \left( \frac{\ell(\ell+1)}{r^2} - \frac{6m(r)}{r^3} + 4\pi \rho(r) - 4\pi P(r) \right).
\end{aligned}
\end{equation}
Considering $\rho(r) > 0$ and  our solutions   that satisfy the DEC  $\rho \ge P$, the effective potentials remain strictly positive-definite throughout the entire domain for $\ell \ge 2$, consistent with the horizonless characteristic of our solutions ($2m(r) < r$). Consequently, the differential operator $\mathcal{D} = -\frac{\partial^2}{\partial r_*^2} + V(r)$ is a positive self-adjoint operator  \cite{Konoplya:2011qq}, ensuring that the axial perturbations is stable.

\section{Conclusions}
In this paper, we have investigated a novel class of non-singular, horizonless compact star solutions sourced by isotropic perfect fluid dark matter halos. By adopting the Einasto and Dehnen galactic density profiles as the  gravitational sources, we have numerically integrated the Einstein field equations and the Tolman-Oppenheimer-Volkoff equation. Our results demonstrate that isotropic dark matter halos can support stable, regular compact objects without the need for the stringent anisotropic pressure. Considering the physical viability of these configurations is intrinsically bounded by the dominant energy condition, we identified critical thresholds for the model parameters—specifically $k_c$, $\rho_{0,c}$, and $\gamma_c$—beyond which the isotropic pressure exceeds the energy density, signaling a transition into non-physical regimes. 

Furthermore, we performed a linear stability analysis against axial gravitational perturbations using the Regge-Wheeler formalism. By restricting our study to the parameter space where the DEC is satisfied, we proved that the corresponding effective potentials remain strictly positive-definite throughout the entire space domain, thereby guaranteeing the robust stability of these dark-matter-sourced compact stars. Our findings suggest that such objects could serve as viable astrophysical candidates for dark matter concentrations in galactic centers, providing a theoretical framework for further studies on their gravitational wave signatures and observational shadows.

\section{Acknowledgment}
This work is supported by the National Natural Science Foundation of China (Grants No.~12275110 and No.~12247101) and National Key Research and Development Program of China (Grant No. 2020YFC2201503).

\end{document}